\documentclass[fleqn,10pt]{wlscirep}
\usepackage[utf8]{inputenc}
\usepackage[T1]{fontenc}
\usepackage{graphicx}
\usepackage{subcaption}
\usepackage{placeins}

\usepackage{cite}

\title{On the Particle Image Overlap in Single Camera Defocusing Approaches}

\author[1,2]{Christian Sax}
\author[1]{Jochen Kriegseis}
\affil[1]{Institute of Fluids Mechanics (ISTM), Karlsruhe Institute of Technology (KIT), Kaiserstraße 10, 76131 Karlsruhe, Germany}
\affil[2]{Karlsruhe Institute of Technology, Institute of Applied and Numerical Mathematics, Karlsruhe, 76133, Germany}

\affil[*]{jochen.kriegseis@kit.edu}

\begin{abstract}
Particle image (PI) overlap presents a significant challenge in single-camera particle tracking and sizing techniques such as Defocusing Particle Tracking Velocimetry (DPTV) and Interferometric Particle Imaging (IPI). In DPTV, overlap obscures PI boundaries, complicating the detection and accurate estimation of PI diameter and center position, which increases uncertainty in the reconstructed particle positions. In IPI, overlap reduces the usable area of the interference pattern, limiting the accuracy of particle size determination. This study introduces a statistical model to quantify PI overlap independently from the optical setup, source density, or PI size. The model assumes uniformly distributed, uniformly sized circular PIs and is validated against experimental data featuring non-uniform sizes and mild astigmatism (up to aspect ratios of 1.66), demonstrating strong agreement. The present study reveals that the seeding density $\mathcal{S}$ serves as a strong universal scaling parameter for PI overlap. Key overlap metrics, including the number of overlaps per PI, the fraction of overlap-free PIs, and the usable PI area are analyzed as functions of the  seeding density $\mathcal{S}$. The results reveal a critical threshold at $\mathcal{S} = 0.25$, where each PI experiences, on average, one overlap. The study provides practical guidance for experimental design by linking overlap metrics to controllable parameters such as source density, defocus length, and aperture diameter. The presented models serve as lookup tools to help experimenters maintain PI overlap within acceptable limits, enabling more reliable and quantitatively robust measurements in DPTV and IPI applications.
\end{abstract}
\begin{document}

\flushbottom
\maketitle
%
%
\thispagestyle{empty}


\section{Introduction}

Particle Tracking Velocimetry (PTV) is a widely used technique for Lagrangian flow measurements by tracking individual particles such as tracers. Advanced methods like tomographic PTV \cite{Nishino1989,Maas1993,Schanz2013,Schanz2016} facilitate three-dimensional, three-component (3D-3C) tracking using multiple cameras. In contrast, defocus-based techniques, such as Astigmatism PTV (APTV) \cite{Cierpka2010,KAO1994}, Defocusing PTV (DPTV) \cite{Willert.1992,Olsen2000}, and holographic PTV \cite{BryanstonCross.1992}, enable 3D-3C tracking with a single camera.
Single-camera techniques such as DPTV are especially useful when investigating flows in scenarios with restricted optical access, where multi-camera techniques cannot be employed. In DPTV, particles are imaged out-of-focus, resulting in circular particle images (PIs) whose diameter increases with distance from the focal plane \cite{Olsen2000,Fuchs.2016}. In DPTV, the in-plane and out-of-plane positions of the particles are determined from the PI center and diameter, respectively. To detect and evaluate PIs in DPTV, methods such as the Hough transform \cite{Atherton.1993,Atherton.1999,Leister.2019}, correlation-based approaches \cite{Barnkob2020,Barnkob.2021}, or object detection algorithms like neural networks \cite{Konig.2020,Cierpka.2019,Dreisbach.2022,Sax2023c} are employed.

A related technique, Interferometric Particle Imaging (IPI), determines particle size by analyzing fringe patterns in defocused particle images (PIs) \cite{Konig.1986,Glover.95}. Initially introduced exclusively for droplet sizing under the name interferometric laser imaging for droplet sizing (ILIDS) \cite{Konig.1986,Glover.95,Rousselle1999}, the method was later extended to bubbles \cite{Kawaguchi.2002,Niwa.2000} and subsequently renamed IPI. This technique is particularly advantageous for sizing small particles within a relatively large field of view, where conventional imaging methods struggle due to limited resolution. In IPI, the fringe frequency scales linearly with particle size \cite{Shen2012}. Typically, either the number of fringes is counted or a fast Fourier transform is applied to extract the fringe frequency from the PI.

\begin{figure}[htbp]
    \centering
    \includegraphics[width=0.8\linewidth]{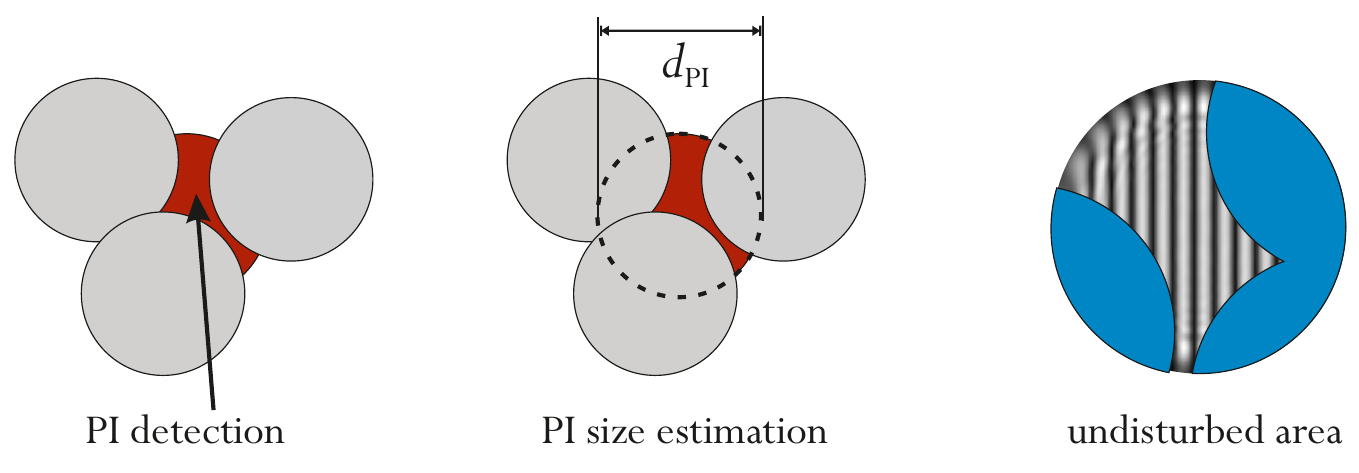}
    \caption{Three issues caused by PI overlap: The detection of covered PIs is more difficult (left). The PI size measurement becomes more challenging and the PI size might be only estimated due to the PI boundaries being covered (middle). The free remaining area (FRA) that can be used in IPI becomes small due to PI overlap (right).}
    \label{fig:Chap6_Overlap_Issues}
\end{figure}

A major concern in both DPTV \cite{Willert.1992,Olsen2000} and IPI \cite{Konig.1986,Glover.95} is particle image (PI) overlap. 
In DPTV, this PI overlap makes identifying individual PIs more difficult, as their boundaries are partially covered by other PIs, compare the left example in Fig.\,\ref{fig:Chap6_Overlap_Issues}. This impairs the detection capability of detection algorithms \cite{Fuchs.2016,Barnkob.2021,Dreisbach.2022,Sax2023c}. PI overlap complicates the measurement of the defocused particle image diameter, $d_\mathrm{PI}$. As a result, fewer particles can be imaged simultaneously to avoid overlap, which in turn reduces the spatial resolution of the reconstructed flow field. As the PI boundary is partially covered in PI overlap, the covered part of the boundary must be assumed based on the curvature of the uncovered part. This leads to an estimation of $d_\mathrm{PI}$ rather than a measurement of $d_\mathrm{PI}$ for some PIs (see middle part of Fig.\,\ref{fig:Chap6_Overlap_Issues}), which in turn increases the uncertainty of the reconstructed $z$-position of the particle. Other approaches simply ignore PIs with overlap to avoid this issue \cite{Fuchs.2016}. Not being able to estimate the PI's boundaries can also affect the determination of the PI's center coordinates ($x_\mathrm{PI},y_\mathrm{PI}$), which can additionally impair subsequent in-plane velocity estimations, from a time series of images. 
The same effects, i.e. limited number of detectable particles and limited $z$-position accuracy, influence the evaluation of particles in IPI. However, in IPI, an additional factor comes into effect. As part of the PI area is covered, only a portion of the interference pattern can be used to size the underlying particle ($d_\mathrm{P}$), compare the right example in Fig.\,\ref{fig:Chap6_Overlap_Issues}. 
Consequently, multiple characteristics are of interest: 
In DPTV, the number of overlaps a PI experiences is of interest, as this provides an estimate of how much of the PI boundary will be covered. 
Simultaneously, the number of PIs in an image that do not experience PI overlap is also of interest, as it offers an estimate of how many particles might be reconstructed with greater accuracy or be used at all.   
Additionally, the maximum amount of overlap a PI experiences is of interest, as it provides information on the size of the boundary segments that are covered, which influences the determination of $d_\mathrm{PI}$.
For IPI, the primary metric of interest is the fraction of the PI area that is covered, or its complement, the remaining free PI area. The remaining assessable PI area is the portion of the PI that can still be used to determine the particle diameter $d_\mathrm{P}$ without the influence of PI overlap.
These metrics are of interest, as evaluation and detection algorithms can be tested for the amount of overlap they can tolerate \cite{Barnkob.2021,Dreisbach.2022,Sax2023c}. 

However, these algorithmic limitations only provide indirect qualitative guidance for an experiment. They do not offer precise recommendations on how much defocusing can be tolerated or how big the aperture diameter can be chosen (i.e. increase of $d_\mathrm{PI}$), or how many particles $N_\mathrm{P}$ can be used in an experiment, to remain within the limits that the post-processing algorithms can handle reliably. 
It is of interest to use larger aperture diameters, as this allows more light to reach the camera chip and consequently, improves the signal-to-noise ratio. More importantly, larger aperture diameters increase the proportionality factor between the particle distance to the focal plane and the PI diameter, effectively enhancing the defocus sensitivity and thereby improving the depth position accuracy. Furthermore, greater defocusing can be advantageous for extending the measurement volume in the $z$-direction. However, both effects increase the PI diameter and, therefore, the likelihood of PI overlap.

Therefore, the objective of the present study is to provide guidance on the extent to which the PI size (i.e. the amount of defocusing and aperture diameter) and the number of PIs can be increased before PI overlap becomes problematic. Furthermore, the trade-off between PI size and source density for constant PI overlap is discussed.
A statistical analysis of PI overlap as a function of PI size $d_\mathrm{PI}$ and the number of particles $N_\mathrm{P}$ is conducted. This analysis allows for the translation of algorithmic limitations into practical experimental settings. 
The theoretical investigation assumes circular PIs, neglecting the influence of astigmatism, and considers uniformly sized PIs. However, testing the model on four different experiments, including cases with non-uniformly sized PIs diameters and slight astigmatism, reveals that the model remains valid despite these simplifying assumptions.

\section{Datasets for the Empirical Investigation}

\subsection{Generation of a Dataset for an Empirical Investigation}

To conduct a statistical investigation of PI overlap, a dataset of images containing PIs is required first. However, experimental images cannot be used for this purpose due to the previously discussed limitations of detecting PIs with high degrees of overlap. Instead, synthetic images are generated in which PIs are randomly distributed, allowing full control over and access to the ground truth positions and sizes of all PIs.

The positions of the PIs in the images were generated under the assumption of a uniform distribution of PI center coordinates, analogous to the commonly assumed uniform particle distribution in Particle Image Velocimetry (PIV). However, this assumption does not hold under all experimental conditions. In certain cases, such as sprays or bubble columns, the spatial distribution of PIs is influenced by the underlying flow structures and is, therefore, non-random. For the purposes of the present study, the uniform distribution serves as a practical and simplifying approximation. Given the wide variety of possible flow topologies, assuming a uniform distribution, simplifies the analysis while still capturing the essential statistical behavior. As will be demonstrated later, the uniform distribution provides a sufficiently accurate approximation for the intended analysis.

\begin{figure*} [htb]
\centering
    \includegraphics[width=0.8\textwidth]{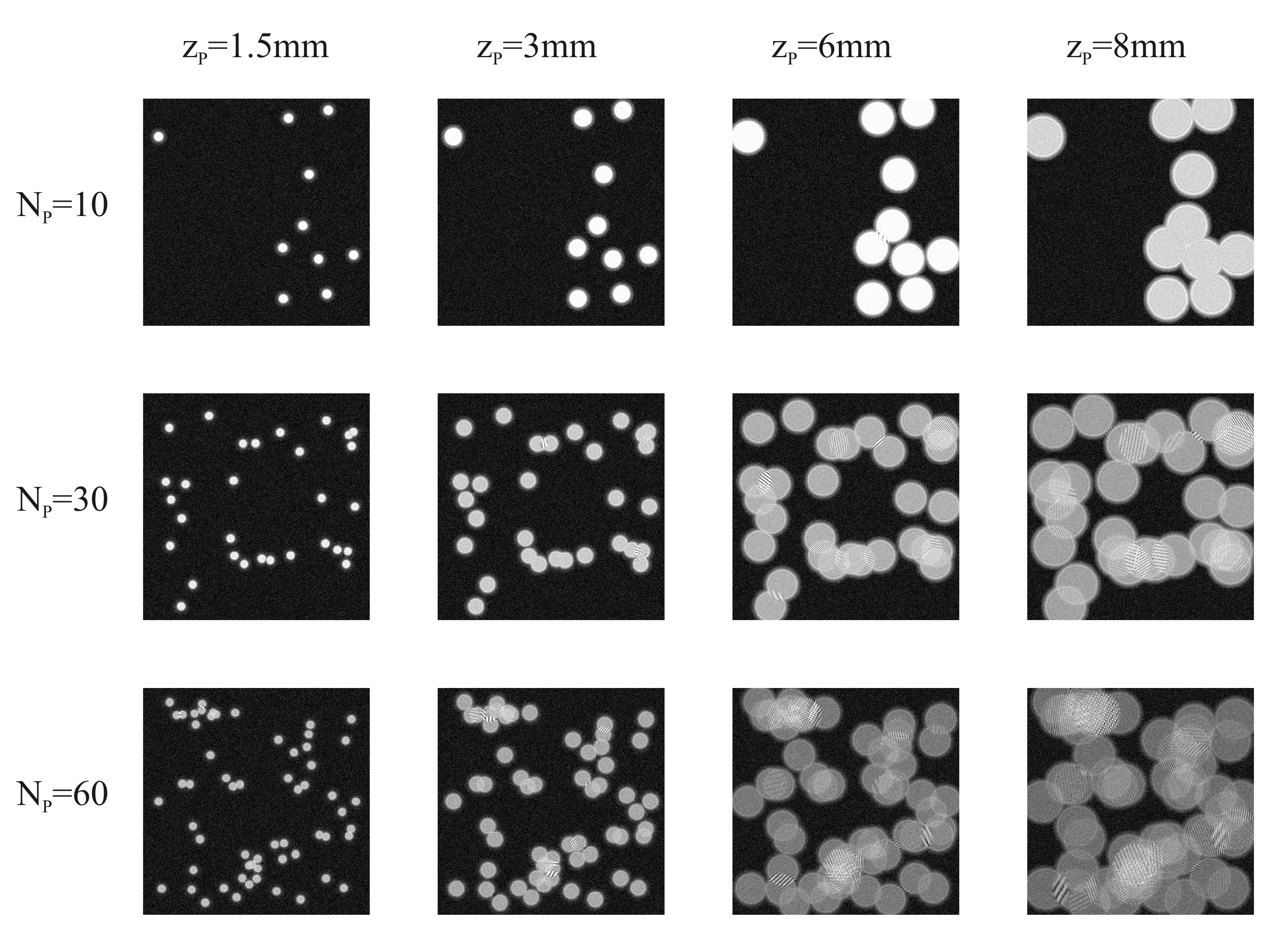}
\caption{Overview over the dataset for the empirical investigation of the PI overlap visualized. Six different number of particles per image $N_\mathrm{P}=[10,20,30,40,50,60]$ and ten different defocus lengths (i.e. PI sizes) $z_\mathrm{P}=[1,1.5,2,3,4,5,6,7,8,9]$\,mm were investigated, resulting in $6\times10$ sets of 400 images each. The images were generated using the physical model from Sax\,\textit{et al.}\,\cite{Sax2025_IP} with parameters from the experiment described in \cite{Sax2023d}.}
\label{fig:Chap6_Defocus_Matrix}
\end{figure*}

To ensure realistic PI sizes relative to the field of view, the optical parameters from the bubble sizing experiment in Sax\,\textit{et al.}\,\cite{Sax2023d} were used to determine the PI diameters. This experiment involves a column of air bubbles ranging from 20-200\,\textmu m in diameter, generated by electrolysis and imaged out-of-focus within a $7\times7$\,mm field of view in a rectangular water tank. Further details can be found in Sax\,\textit{et al.}\,\cite{Sax2023d}.
However, all dimensional quantities were converted into a dimensionless form, making the results applicable to any optical setup.
Both the size and number of PIs were varied in the synthetic images. To reduce the dimensionality of the parameter space and simplify the evaluation, uniformly sized PIs were assumed. This represents scenarios in which particles are located at a single $z$-position, such as in light sheet-based measurements. In cases involving non-uniformly sized PI diameters, the mean PI diameter can be used as a representative value for computing overlap metrics. Although the broader size distribution is expected to increase the variance of these overlap metrics, it is not expected to significantly affect their mean values.
Six different numbers of PIs per image are considered: $N_\mathrm{P} = \{10, 20, 30, 40, 50, 60\}$. The PI diameter is influenced by both the aperture diameter and the defocus length. While the aperture diameter affects the proportionality factor between defocus and PI diameter, defocusing itself shifts all PIs toward larger diameters. However, since both the aperture diameter and the $z$-position (when sufficiently far from the focal plane) have a linear effect on PI size, only the $z$-position was varied in this study to control the PI diameter. In total, ten different defocus lengths are investigated: $z_\mathrm{P} = \{1, 1.5, 2, 3, 4, 5, 6, 7, 8, 9\}$\,mm. The total parameter space thus spans $10 \times 6$ subsets. The PI size corresponding to each defocus distance was calculated using the formula from Shen\,\textit{et al.}\,\cite{Shen2012}, which relies on the linear defocus relationship valid under sufficient defocusing conditions \cite{Fuchs.2016}. Consequently, image generation was reduced to placing a fixed number of circular PIs, each with a computed defocus diameter, at random positions within the image. The structure of the test dataset is illustrated in Fig.\,\ref{fig:Chap6_Defocus_Matrix}.
Each of the $10 \times 6$ subsets contains 400 images to ensure statistical convergence of the evaluated parameters. Convergence validation is shown in Fig.\,\ref{fig:Appendix_Chap6_StatConv} in the Appendix\,\ref{Appendix:Chap6_PI-Overlap}.

\subsection{Experimental Datasets for Validation}

\begin{figure}[htb]
    \centering
    \begin{subfigure}[b]{0.23\linewidth}
        \centering
        \includegraphics[width=\linewidth]{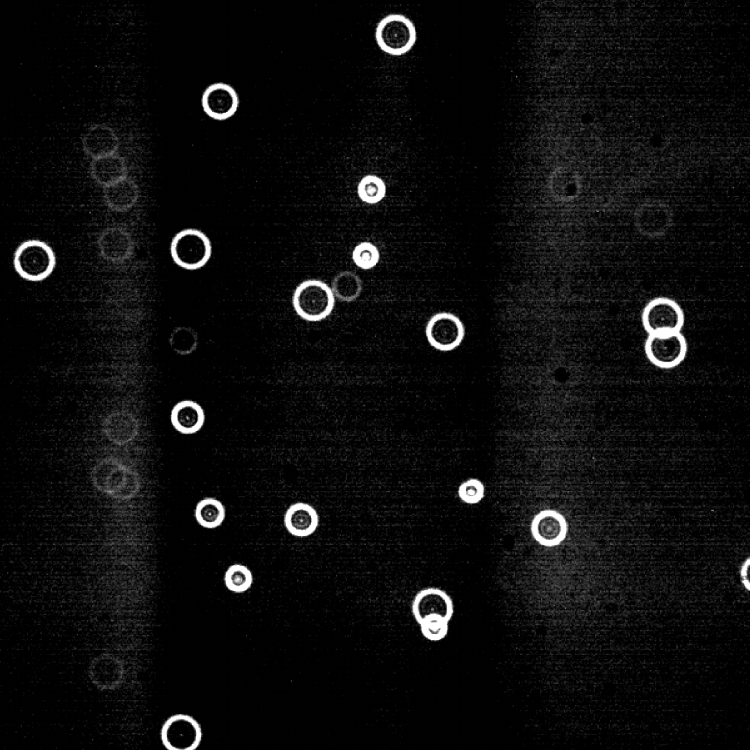}
        \caption{Exp.\,1: Rotating Disk}
        \label{fig:Chap6_Validation_Images-1}
    \end{subfigure}
    \hfill
    \begin{subfigure}[b]{0.23\linewidth}
        \centering
        \includegraphics[width=\linewidth]{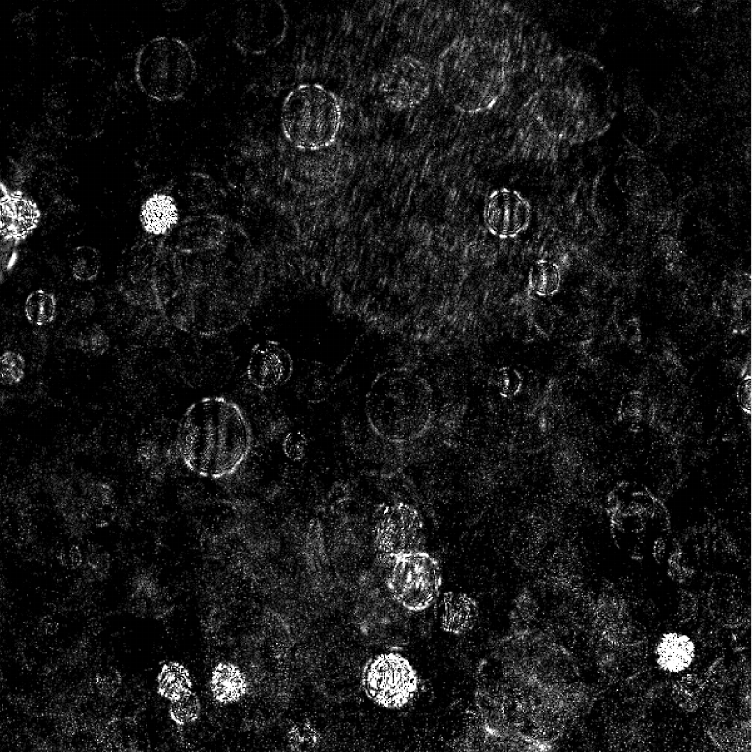}
        \caption{Exp.\,2: Vortex}
        \label{fig:Chap6_Validation_Images-2}
    \end{subfigure}
    \hfill
    \begin{subfigure}[b]{0.23\linewidth}
        \centering
        \includegraphics[width=\linewidth]{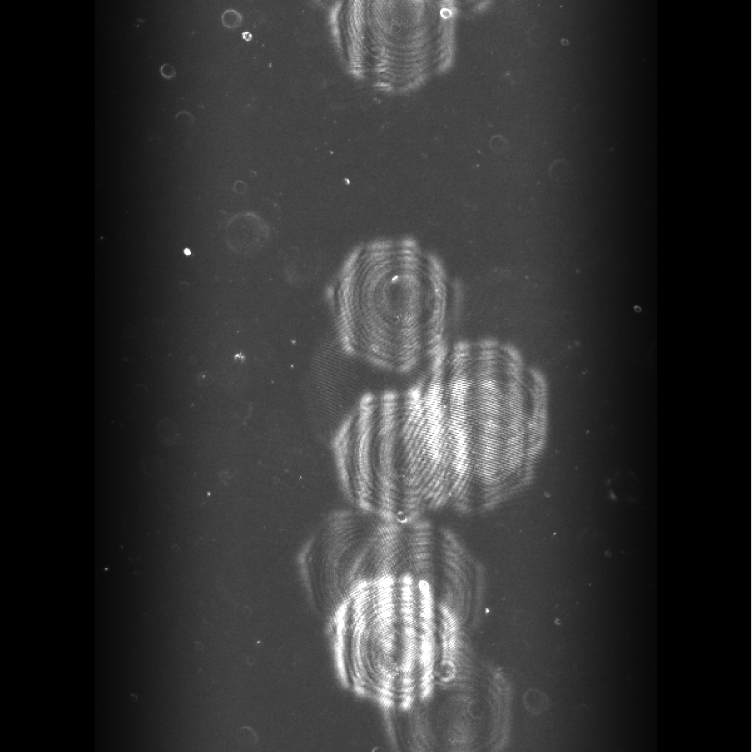}
        \caption{Exp.\,3: Bubble Column}
        \label{fig:Chap6_Validation_Images-3}
    \end{subfigure}
    \hfill
    \begin{subfigure}[b]{0.23\linewidth}
        \centering
        \includegraphics[width=\linewidth]{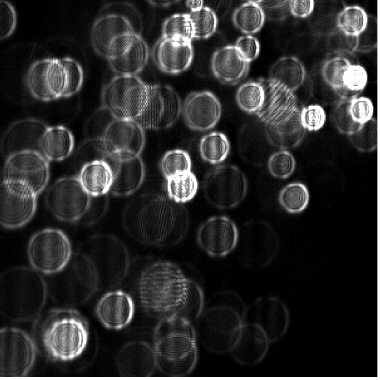}
        \caption{Exp.\,4: Spray}
        \label{fig:Chap6_Validation_Images-4}
    \end{subfigure}
    \caption{Exemplary images from the four different experiments used to validate the PI overlap model. , Exp.\,1\cite{Leister.2021} shows tracers in a gap flow (a), Exp.\,2\cite{LangeSax2024b_conf} shows tracers and bubbles in a vortex like rotating flow (b) and Exp.\,3\cite{Sax2023d} shows shows a bubble column in a resting fluid (c), Exp.\,4\cite{poeppe2023} shows droplets in a spray behind a nozzle (d).}
    \label{fig:Chap6_Validation_Images}
\end{figure}

In the synthetic dataset, two key assumptions were made: (1) PIs are uniformly distributed across the image domain, and (2) all PIs within a given image possess an identical diameter. Although these assumptions may not strictly hold under real experimental conditions, they provide sufficiently accurate approximations, as will be demonstrated through comparison of the theoretical model with the validation experiments.
To evaluate whether the empirical model from the synthetic dataset generalizes to real-world scenarios, four validation image sets were used. These sets originate from four distinct experiments, each involving different flow topologies and optical setups. Representative images from each dataset are shown in Fig.\,\ref{fig:Chap6_Validation_Images}.

The first experiment shows tracer particles in an open wet clutch, which is a gap flow between a rotating disk and a stationary wall. A total of 30 images, like the one in Fig.\,\ref{fig:Chap6_Validation_Images-1}, containing 772 PIs were manually labeled (tracer size $d_\mathrm{P}=9.84\,$\textmu m, PI diameter $d_\mathrm{PI}=[15,31]\,$px). More details on the experiment can be found in Leister\,\textit{et al.}\,\cite{Leister.2021}.  
The second experiment shows the PIs of both tracers (tracer size $d_\mathrm{P}=20\,$\textmu m, PI diameter $d_\mathrm{PI}=[19,273]\,$px) and bubbles (bubble size $d_\mathrm{P}=[20,250]\,$\textmu m, PI diameter $d_\mathrm{PI}=[19,273]\,$px) in a glass cylinder (outer diameter 105\,mm, wall thickness 5.8\,mm). The presence of the glass cylinder introduced mild astigmatism in the captured images, with the maximum observed aspect ratio (width/height) reaching 1.66. A magnetic stirrer was used to create a vortex-like, rotating flow. A total of six images, like the one in Fig.\,\ref{fig:Chap6_Validation_Images-2}, containing 303 PIs were labeled manually. Further information on the experiment can be found in Lange\,\textit{et al.}\,\cite{LangeSax2024b_conf}.
The third experiment shows a column of air bubbles (bubble size $d_\mathrm{P}=[20,250]\,$\textmu m, PI diameter $d_\mathrm{PI}=[110,122]\,$px) in an otherwise stationary fluid (water). A total of 40 images, like the one in Fig.\,\ref{fig:Chap6_Validation_Images-3}, containing 201 PIs were labeled manually. More details on the experiment can be found in Sax\,\textit{et al.}\,\cite{Sax2023d}.
The fourth experiment involves spraying droplets (droplet size $d_\mathrm{P}=[20,60]\,$\textmu m, PI diameter $d_\mathrm{PI}=[16,30]\,$px) from a nozzle outlet. A total of five images, similar to Fig.\,\ref{fig:Chap6_Validation_Images-4}, with a total of 451 PIs, were labeled manually. Details on the experiment can be found in Pöppe\,\cite{poeppe2023}.
These four experiments represent markedly different flow scenarios: Experiments 1 and 2 involve shear and rotating flows with uniformly distributed PIs, whereas Experiments 3 and 4 feature bubble columns and sprays, which exhibit non-uniform PI distributions. By analyzing all four cases, it becomes possible to assess the validity of the uniform distribution assumption across a range of flow conditions.

For each experiment, the images were manually labeled by drawing bounding boxes (BBs) around each PI. The circular representation of each PI was then derived from the center coordinates and the width and height of the corresponding BB. In the case of Exp.\,3, the images exhibit slight astigmatism, and the PIs were approximated as ellipses with principal axes defined by the BB dimensions. This mild astigmatism in Exp.\,3 also provides an opportunity to assess the influence of weak optical distortion on the validity of the results. 

\section{Definition of Overlap Metrics}

To gain insights into the nature of PI overlap from the test dataset, robust measures, independent of used experimental parameters (i.e. non-dimensional), must be defined first for the evaluation. 

\subsection{Metrics for the Number of Overlaps}

To define the number of overlaps a PI $j$ has with other PIs $\ell$ in an image, PI overlap must first be detected. To do this, the intersection-over-union (IoU) measure can be used, compare Fig.\,\ref{fig:Chap6_OverlapMetrics_Critic}. The IoU for the two PIs 
\begin{equation}
    \mathrm{IoU}(\mathrm{PI}_\ell,\mathrm{PI}_j) = \frac{A_{\mathrm{PI},j} \cap A_{\mathrm{PI},\ell}}{A_{\mathrm{PI},j} \cup A_{\mathrm{PI},\ell}}
\end{equation}
can be computed directly from the two circular areas $A_{\mathrm{PI}}$. An overlap between PI $j$ and PI $\ell$ is then detected if the condition $\mathrm{IoU}(\mathrm{PI}_\ell,\mathrm{PI}_j) > 0$ is met. This check is performed for every PI pairing in the image to construct an IoU matrix.
The number of overlaps of $\mathrm{PI}_j$ with other PIs is given by the cardinality of the set 
\begin{equation}
    N_{\mathrm{OL},j} = |\{\mathrm{PI}_\ell\,\,|\,\, \mathrm{IoU}(\mathrm{PI}_\ell,\mathrm{PI}_j)>0 \}|
\end{equation}
of PIs that have an IoU>0 with $\mathrm{PI}_j$. The number of overlaps per PI $N_{\mathrm{OL},j}$ indicates the degree of PI overlap that a given PI $j$ experiences. The degree of the PI overlap, characterizing a cluster of overlapping PI is, therefore, described by $N_{\mathrm{OL},j}$, with a first degree overlap describing the overlap of two PIs, a second degree corresponds to a triplet overlap cluster and so on.
Finally, the fraction of PIs without any overlap, relative to the total number of PIs, $N_\mathrm{noOL}/N_\mathrm{P}$, is also of interest. The number of PIs that do not overlap with any other PIs is defined by the set
\begin{equation}
    N_{\mathrm{noOL}} = |\{\mathrm{PI}_j\,\,|\,\, \mathrm{IoU}(\mathrm{PI}_\ell,\mathrm{PI}_j)=0 \}|.
\end{equation}

\subsection{Metrics for the Amount of Overlaps}
\FloatBarrier

\subsubsection*{Metrics for First Degree Overlaps}
\begin{figure}
    \centering
    \includegraphics[width=0.6\linewidth]{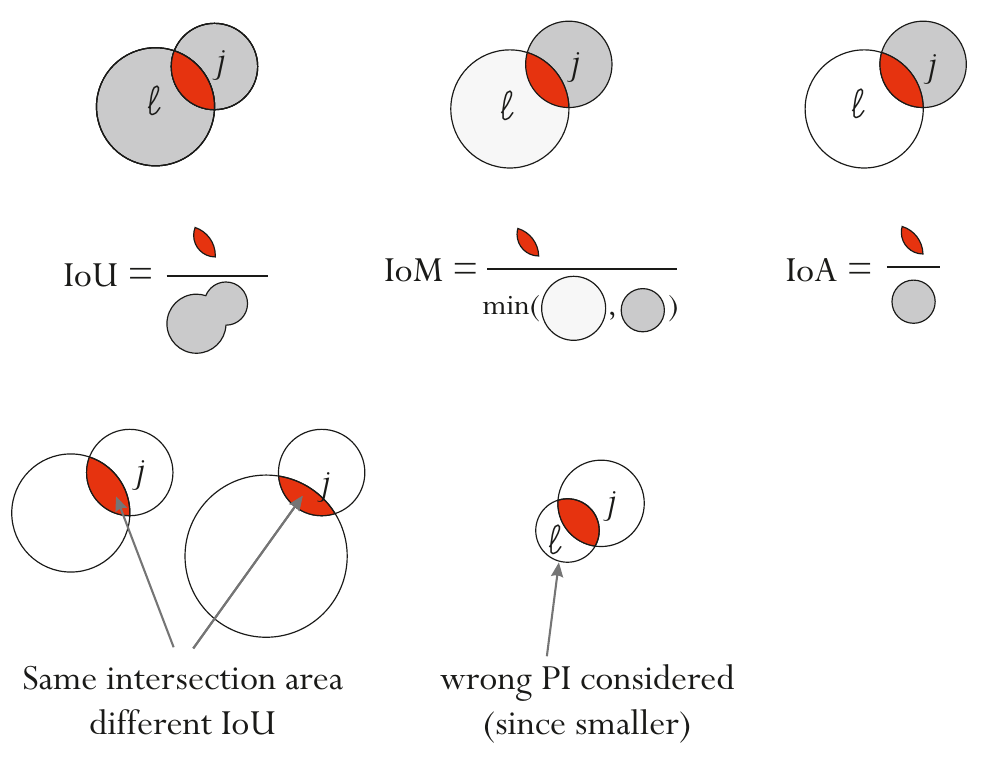}
    \caption{Different measures to describe first degree PI overlaps: IoU (left), IoM (middle) and IoA (right). The issues with the IoU and the IoM are visualized.}
    \label{fig:Chap6_OverlapMetrics_Critic}
\end{figure}

There are many measures to quantify the amount of PI overlap. Dreisbach\,\textit{et al.}\,\cite{Dreisbach.2022} used the IoU to characterize the amount of overlap between two PIs. The IoU is a good measure to describe a first degree overlap, when the overlap in relation to the whole first degree cluster is of interest. However, the IoU has the disadvantage of referencing the intersection area (i.e. the overlapped region) to the union area of the two PIs. This means that for the same intersection area, different IoU values can result depending on the size of the other PI, as illustrated in Fig.\,\ref{fig:Chap6_OverlapMetrics_Critic}. This renders the IoU a non-robust measure when the focus is on the covered area of a specific PI. Moreover, the IoU does not provide direct information about the amount of area overlapped on PI $j$; it only reflects the union area, which also varies with the distance between the PI centres. 

An alternative measure is the Intersection over Minimum (IoM):
\begin{equation}
    \mathrm{IoM}(\mathrm{PI}_\ell,\mathrm{PI}_j) = \frac{A_{\mathrm{PI},j} \cap A_{\mathrm{PI},\ell}}{\mathrm{min}(A_{\mathrm{PI},j},A_{\mathrm{PI},\ell})}
\end{equation}
also known as the Szymkiewicz–Simpson Coefficient \cite{Vijaymeena2016}. This measure was used by Sax\,\textit{et al.}\,\cite{Sax2023c} to characterize PI overlap.  The IoM has the advantage over the IoU of referencing the overlapped area to the area of only one PI, making it more focused on the extent of coverage. However, when two overlapping PIs $j$ and $\ell$ are considered, the reference area changes depending on which PI is smaller, rather than consistently referring to PI $j$, which renders the measure problematic for the use in higher degree overlaps. This effect is illustrated in Fig.\,\ref{fig:Chap6_OverlapMetrics_Critic}.
To address this issue, the IoM is slightly modified. Instead of referencing the minimum area when evaluating the overlap of PI $j$ with other PIs, the area $A_{\mathrm{PI},j}$ of PI $j$ is always used as the reference: 
\begin{equation}
    \mathrm{IoA}(\mathrm{PI}_\ell,\mathrm{PI}_j) = \frac{A_{\mathrm{PI},j} \cap A_{\mathrm{PI},\ell}}{A_{\mathrm{PI},j}}
\end{equation}
The measure is, consequently, called intersection over area (IoA). The IoA has the advantage of consistently referring to the same PI in an PI overlap.
The IoA is equivalent to the IoM for uniformly sized PIs.

The intersection area of two circles of radii $r_{\mathrm{PI}j}$ and $r_{\mathrm{PI}\ell}$ (i.e. $d_\mathrm{PI}/2$) with the Euclidean distance $d_{\mathrm{C},j\ell}$ between the circle centers can be derived from geometrical considerations \cite{WeissteinCircleIntersection}:
\begin{equation}
    A_{\mathrm{PI},j} \cap A_{\mathrm{PI},\ell} = \begin{cases}
        r_{\mathrm{PI}j}^2 \arccos\left(\frac{d_{\mathrm{C},j\ell}^2 + r_{\mathrm{PI}j}^2 - r_{\mathrm{PI}\ell}^2}{2 d_{\mathrm{C},j\ell} r_{\mathrm{PI}j}}\right) +r_{\mathrm{PI}\ell}^2 \arccos\left(\frac{d_{\mathrm{C},j\ell}^2 + r_{\mathrm{PI}\ell}^2 - r_{\mathrm{PI}j}^2}{2 d_{\mathrm{C},j\ell} r_{\mathrm{PI}\ell}}\right) \\
        \quad - \frac{1}{2}\sqrt{(-d_{\mathrm{C},j\ell} + r_{\mathrm{PI}j} + r_{\mathrm{PI}\ell})(d_{\mathrm{C},j\ell} + r_{\mathrm{PI}j} - r_{\mathrm{PI}\ell})} \\
        \qquad\sqrt{(d_{\mathrm{C},j\ell} - r_{\mathrm{PI}j} + r_{\mathrm{PI}\ell})(d_{\mathrm{C},j\ell} + r_{\mathrm{PI}j} + r_{\mathrm{PI}\ell})}, & \text{if } r_{\mathrm{PI}j} + r_{\mathrm{PI}\ell} > d_{\mathrm{C},j\ell} \\
        0, & \text{if } r_{\mathrm{PI}j} + r_{\mathrm{PI}\ell} \leq d_{\mathrm{C},j\ell}
    \end{cases}
\end{equation}

More recently, Xu\,\textit{et al.}\,\cite{XuTropea2024} introduced an alternative measure to characterize PI overlap. They defined the overlap ratio (OLR) of a PI $j$ as:
\begin{equation}
     \mathrm{OLR}=\frac{r_{\mathrm{PI},j}+r_{\mathrm{PI},\ell}-d_{\mathrm{C},j\ell}}{2 \min(r_{\mathrm{PI},j},r_{\mathrm{PI},\ell})}.
\end{equation}
This measure quantifies the difference between the touching distance of the two PIs, given by $r_{\mathrm{PI},j} + r_{\mathrm{PI},\ell}$, and their actual centre-to-centre distance, normalized by the diameter of the smaller PI.
The OLR (with reference to the smaller PI) and the IoA are consistent in the sense that, for different combinations of PI sizes and distances yielding the same IoA, the OLR also remains constant. However, the two measures are not equivalent, i.e., $\mathrm{IoA} \neq \mathrm{OLR}$. While the OLR emphasizes the spatial separation between PIs, the IoA directly quantifies the overlapped area. A limitation of the OLR, similar to that of the IoM, is that it always normalizes with respect to the smaller PI, and thus does not consistently refer to the same PI across comparisons. For these reasons, the IoA is adopted in the present work.

\subsubsection*{Metrics for Higher Degree Overlaps}

\begin{figure}[h]
    \centering
    \includegraphics[width=1\linewidth]{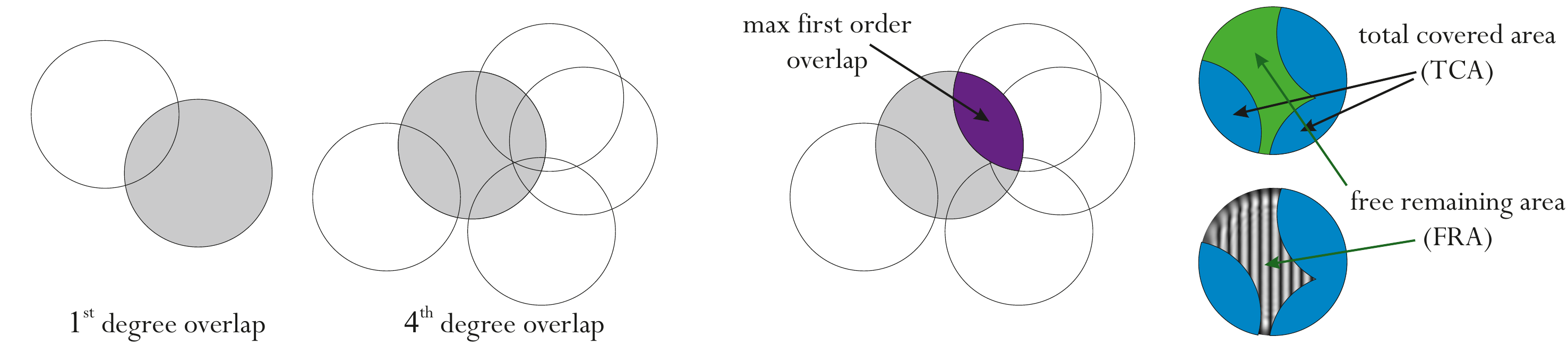}
    \caption{First and higher degree PI overlaps visualized (left). Metrics for higher degree PI overlaps are visualized (right). The FRA and TCA are shown on the example of a PI in IPI.}
    \label{fig:Chap6_OL_Order}
\end{figure}

The IoA can be used to characterize first degree PI overlaps, i.e. the overlap between two PIs. Higher degree overlaps are visualized in Fig.\,\ref{fig:Chap6_OL_Order}. To investigate such higher degree PI overlaps, i.e. between PI $j$ and other PIs $\ell=1,2,..,N_\mathrm{OL}$, further metrics must be defined. 
The $\mathrm{IoA}_\mathrm{max} = \mathrm{max}(\mathrm{IoA}(\mathrm{PI}_{1},\mathrm{PI}_j),..,\mathrm{IoA}(\mathrm{PI}_{N_\mathrm{OL}},\mathrm{PI}_j))$ provides a measure for the largest first degree PI overlap, that the PI $j$ experiences in a higher degree cluster, compare middle part of Fig.\,\ref{fig:Chap6_OL_Order}. 

For IPI, the total fraction of the PI that is covered, or its complement, the remaining free area, is often of interest, as it provides insight into how much of the interference pattern can be used for undisturbed evaluation (see the right example of Fig.\,\ref{fig:Chap6_OL_Order}).

\begin{figure}[h]
    \centering
    \includegraphics[width=0.5\linewidth]{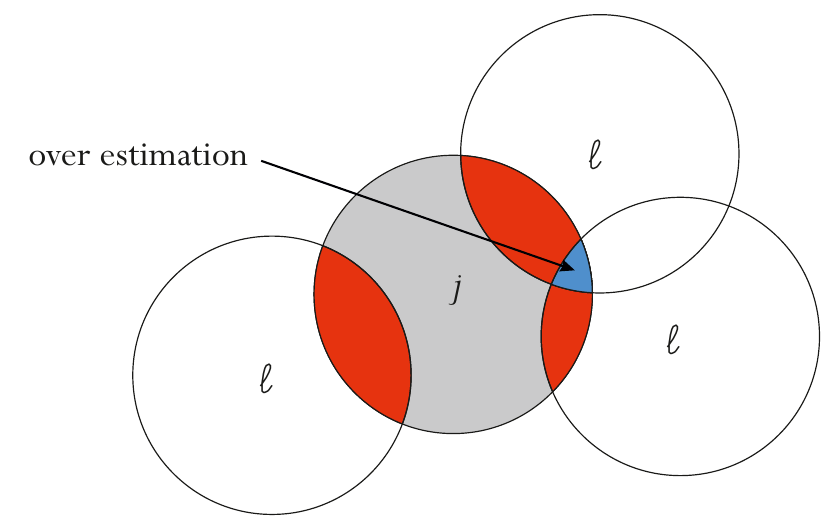}
    \caption{A higher degree PI overlap visualized. The gray area is the PI for which the PI overlap is regarded. The red area is the intersection area, and the blue area shows the twice overlapped area of the PI. The summed IoA overestimated the TCA of the PI by the blue area.}
    \label{fig:Chap6_IoA_Overestimation}
\end{figure}

However, simply summing every IoA of PI $j$ with any other PI $\ell=1,2,..,N_\mathrm{OL}$ overestimates the covered PI area as areas repeatedly overlapped are also counted multiple times, compare Fig.\,\ref{fig:Chap6_IoA_Overestimation}. To determine the free remaining area (FRA), multiple times overlapped areas of the PI should only be considered once.

A robust measure for this case is the total covered area (TCA) and its complement the FRA of PI $j$. The total covered area is defined by the union area of all first degree intersection areas that PI $j$ has with other PIs $\ell$. Instead of using the absolute value of the covered area, the fraction of the covered area over the area of PI $j$ is used 
\begin{equation}
    \mathrm{TCA}_j = \frac{\underset{\ell \neq j}{\bigcup} \left(A_{\mathrm{PI},j} \cap A_{\mathrm{PI},\ell} \right)}{A_{\mathrm{PI},j}}
\end{equation} 
as it places the $\mathrm{TCA}_j\in [0,1]$. The complement FRA can then be simply defined as
\begin{equation}
    \mathrm{FRA}_j = 1-\mathrm{TCA}_j
\end{equation}
the remaining area fraction. The TCA and FRA are visualized in Fig.\,\ref{fig:Chap6_OL_Order}.

To compute the TCA, the union of the intersection areas mus be known. This can be computed from the $\mathrm{IoA}$ in principle using the inclusion-exclusion principle of Poincaré and Sylvester \cite{Henze2011}. However, the computation requires the intersection areas between all combinations of PIs from the first to the highest degree overlap. Since this is computationally unfeasible, a quadrature-like approach is used to compute the TCA in the following.

\subsection{Direct Computation of Overlapped Particle Image Fraction}

\begin{figure}[htbp]
    \centering
    \includegraphics[width=1\linewidth]{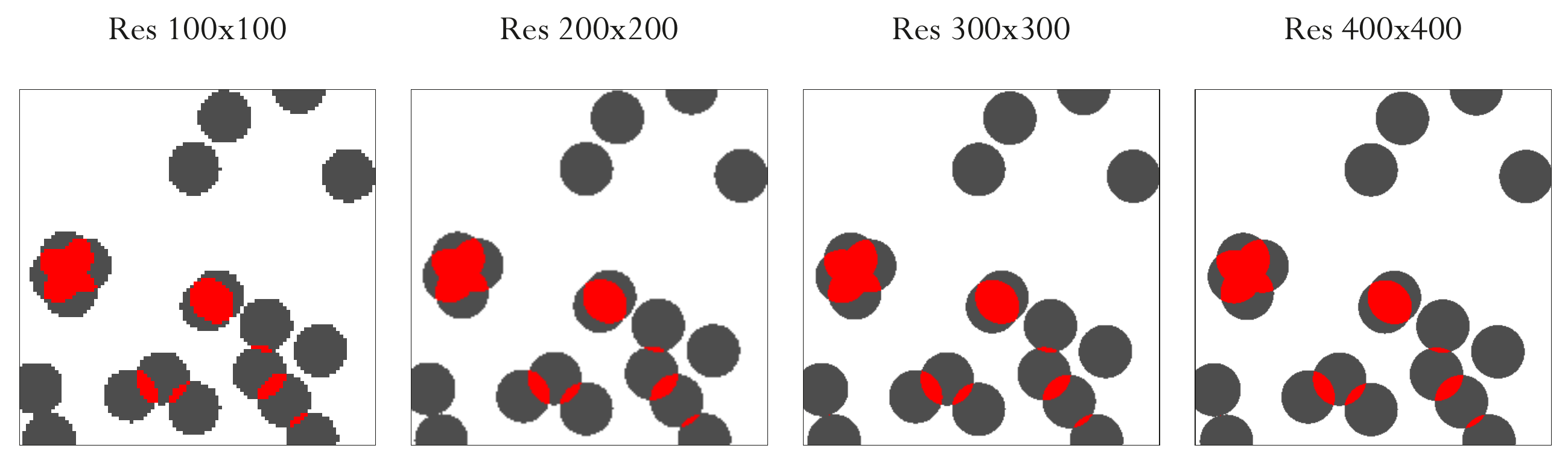}
    \caption{Deterministic pixel counting approach by simulating discretized circles. A circle consists of pixels with a value of one (marked as gray). The overlap areas of each circle are obtained from the number of pixels with a value of larger than one (marked as red). Different grid resolutions of the simulation are shown from 100\,px to 400\,px.}
    \label{fig:Chap6_MonteCarloMethod}
\end{figure}

The TCA is computed using a quadrature-like approach, in which the circle area is discretized into pixels on a Cartesian grid (see Fig.\,\ref{fig:Chap6_MonteCarloMethod}). The circle areas are approximated by pixel counting, analogous to a Riemann sum.
The grid consists of pixels $(u,v)\in \mathbb{Z}^2$.
In this method, each PI is defined by its in-plane position $(x_\mathrm{PI}, y_\mathrm{PI})$ and diameter $d_\mathrm{PI}$, and is represented as a binary circle mask 
\begin{equation}
    M_\mathrm{circle}(u,v) = 
    \begin{cases}
    1, & \text{if } (u - x_\mathrm{PI})^2 + (v - y_\mathrm{PI})^2 \leq (\frac{d_\mathrm{PI}}{2})^2 \\
    0, & \text{otherwise}
\end{cases}
\end{equation}
where the mask takes the value 1 for pixels inside the circle and 0 elsewhere.
By summing these binary masks over a zero-initialized image, a combined mask is formed in which each pixel value indicates the number of overlapping circles. The TCA for a given PI $j$ is approximated by counting the number of pixels within its own circle mask that have a value greater than one, and dividing this by the total number of pixels in that mask.

The set of pixels belonging to PI $j$ is defined as:
\begin{equation}
    M_\mathrm{PI}=\left\{(u,v)\in\mathbb{Z}^2\,\,\middle|\,\,(u - x_\mathrm{PI})^2 + (v - y_\mathrm{PI})^2 \leq (\frac{d_\mathrm{PI}}{2})^2\right\}.
\end{equation}
The TCA is then approximated by
\begin{equation}
    \mathrm{TCA}_j \approx \frac{\underset{u,v\in M_{\mathrm{PI},j}}{\sum} \mathbf{1}(M_{\mathrm{comb}}(u,v)>1)}{|M_{\mathrm{PI},j}|}
\end{equation}
where $\mathbf{1}$ is the indicator function, equal to 1 if the condition is true and 0 otherwise, and 
\begin{equation}
    M_{\mathrm{comb}}(u,v) = \sum_{\ell=1}^{N_\mathrm{P}}M_{\mathrm{circle},\ell}(u,v) 
\end{equation}
is the combined mask of all $N_\mathrm{P}$ PIs.

\begin{figure}[htbp]
    \centering
        \includegraphics[width=0.4\linewidth]{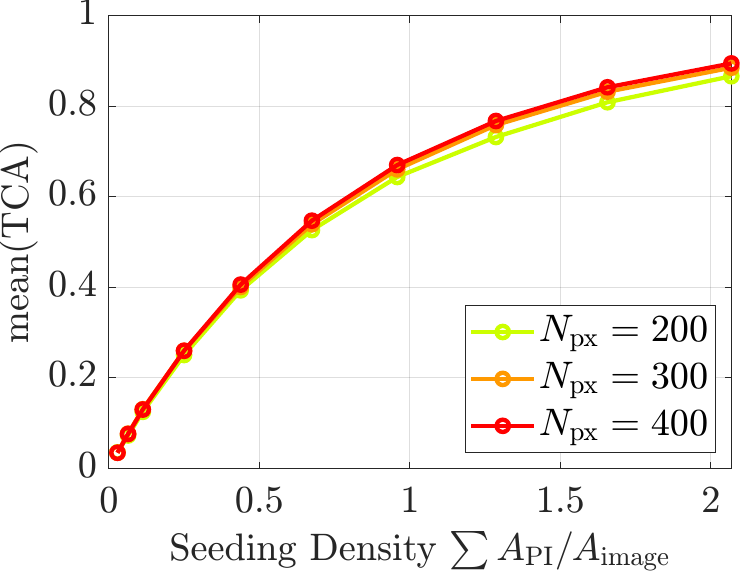}
    \caption{The approximation of the TCA by the quadrature-like method for $N_\mathrm{P}=60$. The discretized simulations are shown for different grid resolutions $N_\mathrm{px}=\{200,300,400\}$. It can be seen that almost no difference between the $N_\mathrm{px}=300$ and $N_\mathrm{px}=400$ resolution can be observed.}
    \label{fig:Chap6_GridConvergence Proof}
\end{figure}

This circle mask method enables direct approximation of the TCA and FRA. However, since the circle areas are discretized onto a pixel grid, the method inherently introduces a discretization error. As it follows a quadrature-like approach, increasing the grid resolution improves the accuracy of the approximation, and the error asymptotically vanishes with finer discretization.
To ensure that the chosen grid resolution was sufficiently fine, the method was executed four times using different grid sizes $(u,v) \in \{1, \ldots, N_\mathrm{px}\} \times \{1, \ldots, N_\mathrm{px}\}$, with $N_\mathrm{px} = \{100, 200, 300, 400$\}; see Fig.\,\ref{fig:Chap6_MonteCarloMethod}. 
Figure~\ref{fig:Chap6_GridConvergence Proof} shows the TCA computed for $N_\mathrm{P} = 60$ PIs as a function of the summed PI area relative to the image area. A slight difference is observed between the 200\,px and 300\,px resolutions, while almost no difference is observed between 300\,px and 400\,px; see Fig.~\ref{fig:Chap6_GridConvergence Proof}. This indicates that a resolution of $N_\mathrm{px} = 400$ is likely sufficient to obtain a reliable approximation of the TCA.

\subsection{Definition of a Scaling Metric}

Before evaluating the overlap metrics as a function of the number of PIs and PI size, a scaling metric that consolidates both $N_\mathrm{P}$ and $z_\mathrm{P}$ into a single value is introduced. 
Cierpka\,\textit{et al.}\,\cite{Cierpka.2011} introduced a modified definition of seeding density $\mathcal{S}$, as a ratio of the total PI area $A_\mathrm{PI}$ to the total image area $A_\mathrm{image}$. This definition
\begin{equation}
    \mathcal{S} = \sum_{\ell=1}^{N_\mathrm{P}} \frac{A_{\mathrm{PI},\ell}}{A_\mathrm{image}}
    \label{Equ:Chap6_SeedingDens_Org}
\end{equation}
is also employed in the present work to summarize both the number and size of PIs. 
The seeding density defined in Eq.\,\eqref{Equ:Chap6_SeedingDens_Org} relates the PI area to the image area, which is determined by the region of interest (RoI) of the respective experiment. Since the RoI is generally unknown, it is here defined as the smallest enclosing rectangle (reflecting the typical rectangular shape of camera sensors) that contains all PIs $\ell$ across all images, i.e., across all time steps $t$. The reference area is thus defined as:
\begin{multline}
    A_\mathrm{image} := \left[\max\left(x_{\mathrm{PI},\ell,t}+\frac{d_{\mathrm{PI},\ell,t}}{2}\right)-\min\left(x_{\mathrm{PI},\ell,t}-\frac{d_{\mathrm{PI},\ell,t}}{2}\right)\right] \left[\max\left(y_{\mathrm{PI},\ell,t}+\frac{d_{\mathrm{PI},\ell,t}}{2}\right)-\min\left(y_{\mathrm{PI},\ell,t}-\frac{d_{\mathrm{PI},\ell,t}}{2}\right)\right]
\end{multline}
The seeding density $\mathcal{S}$ depends 
\begin{equation}
    \mathcal{S} \thicksim N_\mathrm{P}d_\mathrm{PI}^2 \thicksim N_\mathrm{P}z_\mathrm{P}^2
    \label{Equ:Chap6_SeedingDensity_Sub}
\end{equation}
linearly on the number of PIs, $N_\mathrm{P}$, and quadratically on the PI diameter, $d_\mathrm{PI}$. Given the linear relationship between the distance of the PI to the focal plane, $z_\mathrm{P}$, and $d_\mathrm{PI}$ under sufficient defocusing \cite{Fuchs.2016,Olsen2000}, $\mathcal{S}$ is therefore also quadratically dependent on $z_\mathrm{P}$. The same quadratic dependence applies to the aperture diameter, i.e., $\mathcal{S} \thicksim D_\mathrm{a}^2$.

Consequently, $\mathcal{S}$ substitutes the two dimensional parameter space spanned by $N_\mathrm{P}$ and $z_\mathrm{P}$ to a one-dimensional parameter that is independent of physical units. From Eq.~\eqref{Equ:Chap6_SeedingDensity_Sub}, it follows that various combinations of $N_\mathrm{P}$ and $z_\mathrm{P}$ can yield the same seeding density. For instance, an increased number of PIs in the image requires a reduction in defocus by the square root to maintain the same seeding density. Conversely, increased defocus must be compensated by a quadratic reduction in the number of PIs. 
For an arbitrary scaling $\epsilon\in\mathbb{R}$, this relationship between $N_\mathrm{P}$ and $z_\mathrm{P}$ can be expressed as $N_\mathrm{P}z_\mathrm{P}^2 = \epsilon N_\mathrm{P}(z_\mathrm{P}/\sqrt{\epsilon})^2$.
By computing the defined metrics over $\mathcal{S}$, the experimentalist can select an appropriate combination of $N_\mathrm{P}$ and $z_\mathrm{P}$ for a given seeding density.

\section{Scaling Model for Particle Image Overlap}

The means of all previously introduced measures, $N_{\mathrm{OL},j}$, $N_\mathrm{noOL}$, $\mathrm{IoA}_\mathrm{max}$, TCA, and FRA, are computed for the $10 \times 6$ parameter space of the test dataset. This way the influence of PI size and number of PIs on the PI overlap can be investigated. Additionally, are the measures computed for the validation image sets.\\

\begin{figure}[htb]
    \centering
     \begin{subfigure}[b]{0.45\linewidth}
        \centering
        \includegraphics[width=\linewidth]{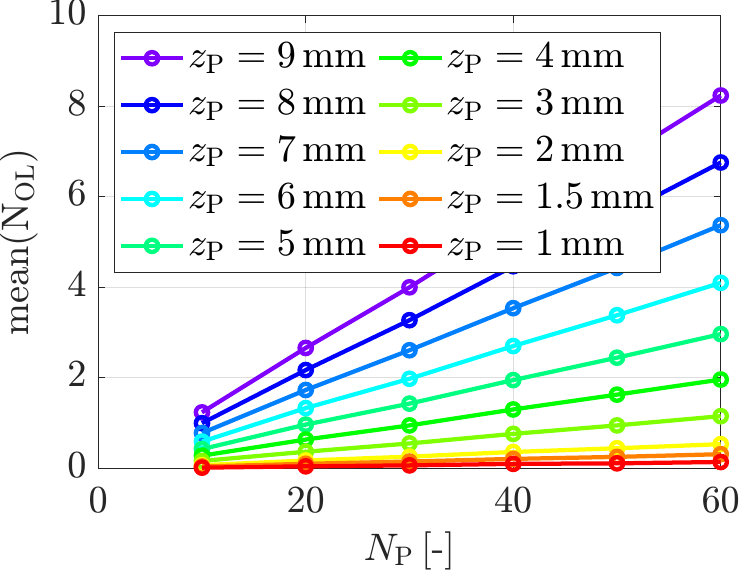}
        \caption{Number over overlap per PI over $N_\mathrm{P}$}
        \label{fig:Chap6_NumOl_NZ-a}
    \end{subfigure}
    \hfill
    \begin{subfigure}[b]{0.45\linewidth}
        \centering
        \includegraphics[width=\linewidth]{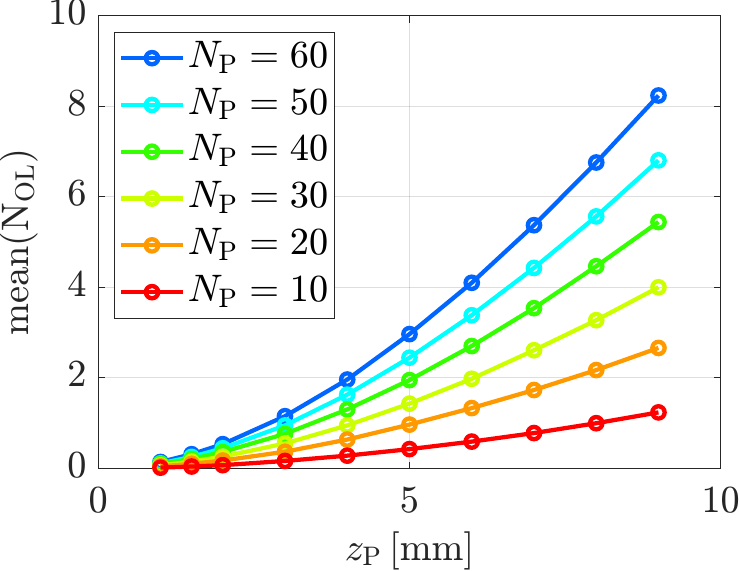}
        \caption{Number over overlap per PI over $z_\mathrm{P}$}
        \label{fig:Chap6_NumOl_NZ-b}
    \end{subfigure}
    \caption{The mean number of overlaps per PI over the number of PIs for different defocus distances (a) and over the defocus distance for different number of PIs (b).}
    \label{fig:Chap6_NumOl_NZ}
\end{figure}

An exemplary plot of the number of overlaps per PI, $N_\mathrm{OL}$, as a function of $N_\mathrm{P}$ and $z_\mathrm{P}$ is shown in Figs.\,\ref{fig:Chap6_NumOl_NZ-a} and \ref{fig:Chap6_NumOl_NZ-b}, respectively. It can be seen that the number of overlaps scales linearly with $N_\mathrm{P}$ and quadratically with $z_\mathrm{P}$. However, the proportional scaling of $N_\mathrm{OL}$ with respect to $N_\mathrm{P}$ varies depending on the $z$-position, and vice versa. While this allows for an assessment of the general behavior of this overlap measure in relation to the experimental parameters, it does not provide a universal scaling law.
In contrast, Fig.\,\ref{fig:Chap6_NumOl-a} presents the number of overlaps per PI plotted against the seeding density $\mathcal{S}$, which combines both parameters $N_\mathrm{P}$ and $z_\mathrm{P}$ into a single metric. It becomes immediately apparent that, when scaled by $\mathcal{S}$, the individual curves collapse onto a single master curve.
As illustrated in Figs.\,\ref{fig:Chap6_NumOl} and \ref{fig:Chap6_TCA}, this behavior is consistent across all previously defined measures.
This observation suggests that $\mathcal{S}$ acts as a robust scaling parameter for PI overlap, indicating a scaling law that is independent of the specific measurement setup.

The second observation that can be drawn is that, the four experiments show a good alignment with the theoretical model, as the data points follow the metric curves closely, compare Figs.\ref{fig:Chap6_NumOl} and \ref{fig:Chap6_TCA}.
This shows that despite all four image sets stemming from experiments with different flow topology, the assumption of uniform distribution describes the distribution of PIs in the image sufficiently well. It also shows that despite using equally sized PIs, the results can be extended to the PI of various sizes from the four experiments. 

The third observation that can be drawn is that the scaling laws also capture the PI overlap behavior observed in Exp.\,3, which exhibits a small amount of astigmatism. It can therefore be assumed that the results also hold for weak astigmatism (aspect ratio\,<\,1.66) in the PIs. However, no definitive statement for stronger astigmatism can be made. 

\subsubsection*{Number Overlaps}

\begin{figure}[htb!]
    \centering
    \begin{subfigure}[b]{1\linewidth}
    \centering
        \includegraphics[width=0.8\linewidth]{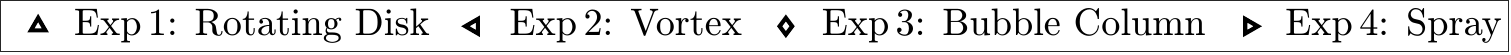}
    \end{subfigure}
    \vskip 0.8cm
    \begin{subfigure}[b]{0.45\linewidth}
        \centering
        \includegraphics[width=\linewidth]{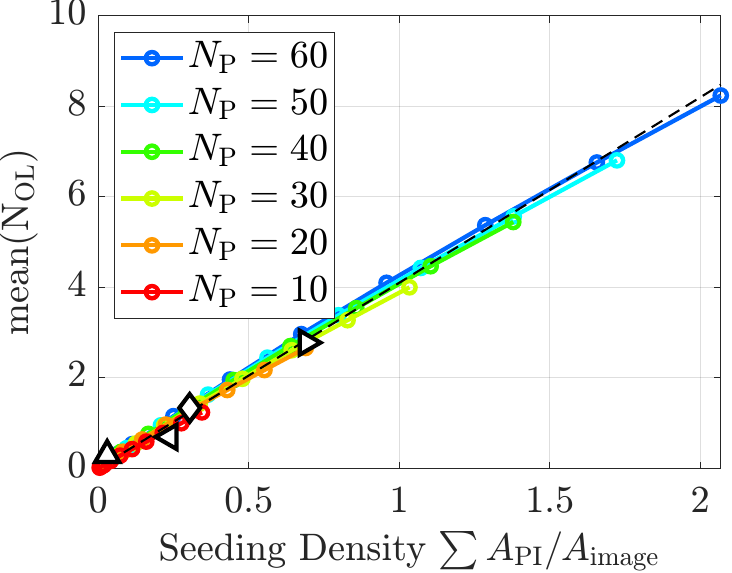}
        \caption{Number over overlap per PI}
        \label{fig:Chap6_NumOl-a}
    \end{subfigure}
    \hfill
    \begin{subfigure}[b]{0.45\linewidth}
        \centering
        \includegraphics[width=\linewidth]{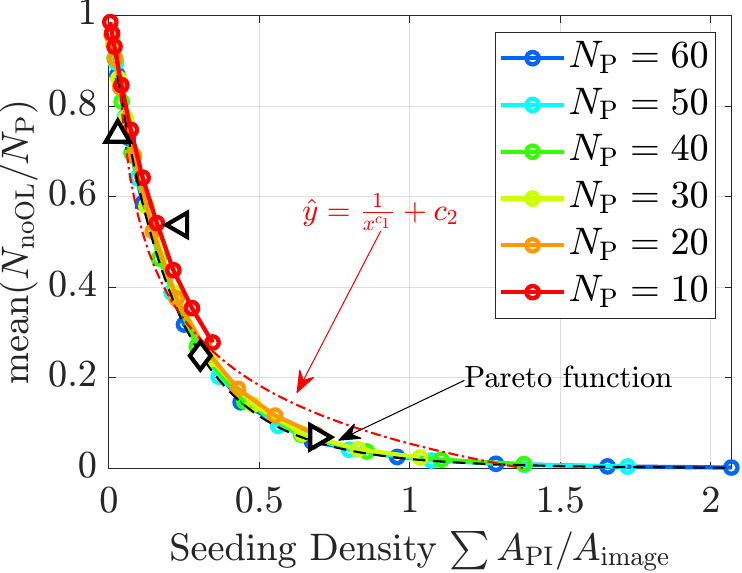}
        \caption{Fraction of PIs without overlap}
        \label{fig:Chap6_NumOl-b}
    \end{subfigure}
    \caption{The mean number of overlaps per PI over the seeding density $\mathcal{S}\thicksim N_\mathrm{P}z_\mathrm{P}^2$ for different number of PIs in the image (a). The mean fraction of PIs without overlap in an image over the seeding density $\mathcal{S}$ and for different number of PIs in the image (b). Shown are also the fits to the curves (\textit{dashed line}\,\, \rule[0.5ex]{0.1cm}{0.5pt}\hspace{0.05cm}\rule[0.5ex]{0.1cm}{0.5pt}\hspace{0.05cm}\rule[0.5ex]{0.1cm}{0.5pt}), which are specified in Tab.\ref{tab:Chap6_TabFits}. The values obtained from the validation experiments are provided by the markers ($ \triangle,\lhd,\Diamond,\rhd$). Images from the validation experiments are shown in Fig.\,\ref{fig:Chap6_Validation_Images}.}
    \label{fig:Chap6_NumOl}
\end{figure}

Fig.\,\ref{fig:Chap6_NumOl-a} shows the mean number of overlaps per PI $j$ (i.e. the degree of PI overlap) in an image as a function of the seeding density, $\mathcal{S}$. It can be observed that the expected number of overlaps per PI, $N_{\mathrm{OL},j}$, increases linearly with $\mathcal{S}$. The trend follows a straight line with a slope of four, indicating that at a critical seeding density of $\mathcal{S} = 0.25$, an average of one overlap per PI can be expected. In other words: the average PI cluster contains two PIs, or the average PI overlap is of degree one. 

Fig.\,\ref{fig:Chap6_NumOl-b} shows the mean fraction of PIs in the image that experience no overlap. This measure is particularly interesting, as it provides insight into the proportion of PIs that can be processed more reliably. The number of overlap-free PIs $N_\mathrm{noOL}$ decreases rapidly for small values of $\mathcal{S}$, with the slope gradually leveling off. This is a significant finding, as it demonstrates that even slight defocusing near the focal plane has a substantial impact on PI overlap, whereas increasing the seeding density beyond $\mathcal{S} = 0.75$ has minimal effect on the number of overlap-free PIs. 
Additionally, Fig.\,\ref{fig:Chap6_NumOl-b} shows that the decline of the fraction of overlap-free PIs follows a Pareto function.

\FloatBarrier

\subsubsection*{First Degree Overlaps Characteristics}

Most detection algorithms are capable of distinguishing PIs involved in multiple small first-degree overlaps, but typically struggle with larger first-degree overlaps. The total overlapped PI area within a cluster does not indicate whether the area is covered by several small or a few large first-degree overlaps. Most studies focusing on PI detection define overlap metrics exclusively for first-degree PI overlaps \cite{Dreisbach.2022,Sax2023c,XuTropea2024,RaoTropea2024}. Therefore, when PI edges are of interest, such as in PI detection for DPTV, it is useful to consider first-degree PI overlap metrics in conjunction with higher-degree metrics.
The maximum IoA is a particularly insightful metric, as it reflects the strongest first-degree overlap within a cluster. When combined with the TCA, it provides insight into how individual PI overlaps contribute to the total covered area, i.e., whether a PI is primarily covered by a single large overlap or by multiple smaller ones.

Fig.\,\ref{fig:Chap6_TCA-a} shows the change of the mean $\mathrm{IoA}_\mathrm{max}$ per PI with $\mathcal{S}$. For low seeding densities a steep incline of the max IoA can be observed, which gradually becomes less steep with increasing $\mathcal{S}$.
It can be seen that the increase in max IoA is approximately linear for $\mathcal{S}<0.25$ with a slope of approx one.

\FloatBarrier
\subsubsection*{Free Particle Area in Higher Degree PI Overlaps}

\begin{figure}[htbp]
    \centering
    \begin{subfigure}[b]{1\linewidth}
    \centering
        \includegraphics[width=0.8\linewidth]{Legend_Only_Exp.pdf}
    \end{subfigure}
    \vskip 0.8cm
    \begin{subfigure}[b]{0.45\linewidth}
        \centering
        \includegraphics[width=\linewidth]{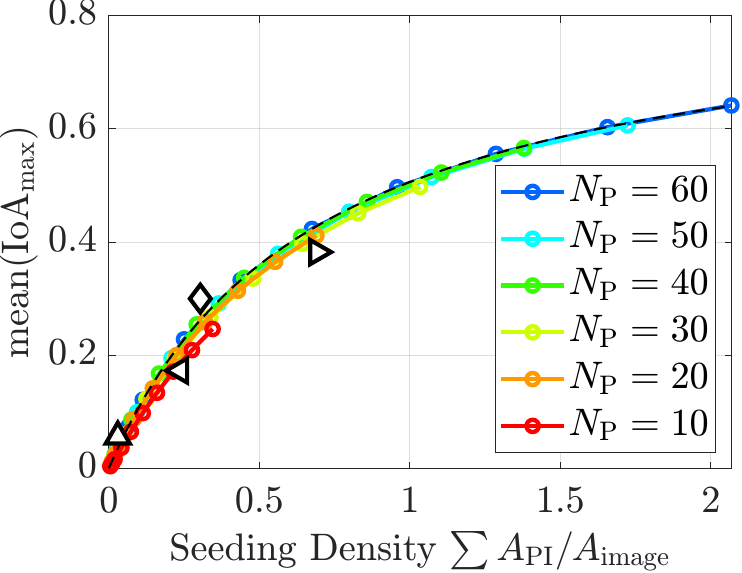}
        \caption{Resolution in TCA Approximation}
        \label{fig:Chap6_TCA-a}
    \end{subfigure}
    \hfill
    \begin{subfigure}[b]{0.45\linewidth}
        \centering
        \includegraphics[width=\linewidth]{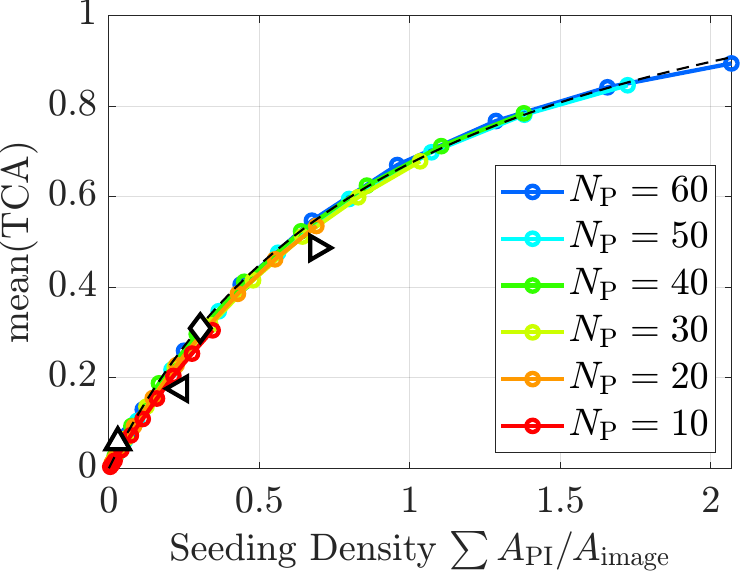}
        \caption{Total Covered Area of a PI}
        \label{fig:Chap6_TCA-b}
    \end{subfigure}
    \caption{The max IoA per PI over the seeding density $\mathcal{S}\thicksim N_\mathrm{P}z_\mathrm{P}^2$ for different number of PIs in the image (a). The mean TCA with $N_\mathrm{px}=400$ over the seeding density $\mathcal{S}$ for different number of PIs in the image (b). Shown is also the fit to the curves (\textit{dashed line}\,\, \rule[0.5ex]{0.1cm}{0.5pt}\hspace{0.05cm}\rule[0.5ex]{0.1cm}{0.5pt}\hspace{0.05cm}\rule[0.5ex]{0.1cm}{0.5pt}), which is specified in Tab.\ref{tab:Chap6_TabFits}. The values obtained from the validation experiments are provided by the markers ($ \triangle,\lhd,\Diamond,\rhd$). Images from the validation experiments are shown in Fig.\,\ref{fig:Chap6_Validation_Images}.}
    \label{fig:Chap6_TCA}
\end{figure}

The extent of overlap in higher-degree PI interactions is described by the TCA and FRA, and constitutes a particularly important measure for IPI, where the PI area is used rather than the edges.
Fig.\,\ref{fig:Chap6_TCA-b} shows how the TCA scales with the seeding density $\mathcal{S}$. The TCA increases steeply at low seeding densities and asymptotically approaches one. Most practical applications will not involve seeding densities greater than $\mathcal{S} = 1$, and thus will not encounter mean TCAs exceeding approximately 70\%. In fact, in many cases, PI overlap should be minimized as much as possible. Therefore, if an average TCA of e.g. 10\% is considered acceptable, the seeding density should be kept below $\mathcal{S} = 0.09$.

The goal of the present work is to provide practical guidance for experimenters on selecting the number of particles and defocus lengths, to ensure that specific limits on PI overlap are maintained. The fit functions for all evaluated measures—obtained via least squares regression—are presented in Tab.\,\ref{tab:Chap6_TabFits}. These functions enable the creation of simple plots for each measure, so that experimenters can use the PI overlap model without the need to repeat the empirical investigation.

\begin{table}[htbp]
    \centering
    \caption{The different metrics to describe overlap and their respective models are shown. The model type as well as the fit function and the parameters are given. The goodness of fit is measured by R-squared.}
    \label{tab:Chap6_TabFits}
    \begin{tabular}{c|c|c|c|c}
       Metric & Model & Model-function & Parameters & R-squared\\ \toprule
       mean($N_\mathrm{noOL}/N_\mathrm{P}$) & Pareto (PDF) & $\mathrm{mean}(N_\mathrm{noOL}/N_\mathrm{P})=\frac{1}{c_1}(1+c_2\frac{(\mathcal{S}-c_3)}{c_1})^{(-\frac{1}{c_2}-1)} $ & $c_1=0.193$ & 99.999\%\\
       & & & $c_2=0.156$ &\\
       & & & $c_3=-0.307$ &\\ \midrule
       $\mathrm{mean}(N_\mathrm{OL})$ & linear & $\mathrm{mean}(N_\mathrm{OL})= c_1\mathcal{S}$ & $c_1=4.063$ & 99.840\%\\ \midrule
       $\mathrm{mean}(\mathrm{IoA}_\mathrm{max})$ & rational function & $\mathrm{mean}(\mathrm{IoA}_\mathrm{max})=\frac{c_1\mathcal{S}}{1+c_2\mathcal{S}}$ & $c_1=1.245$ & 99.999\%\\
       & & & $c_2=1.460$ & \\\midrule
       mean(TCA) & rational function & $\mathrm{mean}(\mathrm{TCA})=\frac{c_1\mathcal{S}}{1+\mathcal{S}}$ & $c_1=1.324$ & 99.956\%\\\bottomrule
    \end{tabular}
\end{table}

\FloatBarrier
\section{Concluding Remarks on Particle Image Overlap}

PI overlap presents a major challenge in DPTV, where incomplete PI boundary information increases detection miss rates and uncertainty in $z$-position estimation. In IPI, it, furthermore, reduces the usable PI area for fringe pattern analysis, which is essential for accurately determining the particle diameter ($d_\mathrm{P}$).
The developed model, which describes PI overlap independently of the optical setup, particle number, or PI size, is based on the assumption of a uniform spatial distribution of PIs with uniform sizes. Despite these simplifying assumptions, a comparison with experimental data, featuring non-uniform particle distributions and multiple PI sizes, shows that the model captures the observed behavior with sufficient accuracy.
The model also performs well under conditions of mild astigmatism (i.e. aspect ratios of 1.66 or smaller). However, its validity under stronger astigmatic distortions remains uncertain and may require further investigation.
The empirical study of PI overlap demonstrates that the seeding density $\mathcal{S}\thicksim N_\mathrm{P}z_\mathrm{P}^2$ serves as a powerful scaling parameter for describing PI overlap independently of the optical system. It was further observed that the PI overlap remains constant for any combination of $N_\mathrm{P}$ and $z_\mathrm{P}$ that results in a constant seeding density. This implies that, for example, the number of particles can be increased linearly while the amount of defocusing is reduced by the square root, without altering the mean PI overlap. 

The number of overlaps experienced by a PI, scales linearly with the seeding density, with a critical threshold at $\mathcal{S} = 0.25$, where an average of one overlap per PI can be expected. This means that if the PIs cover 25\% of the recorded image, the average overlap cluster degree is one. Small variations in defocus length near the focal plane have a significant impact on the fraction of overlap-free PIs, particularly because $\mathcal{S} \thicksim z_\mathrm{P}^2$. At approximately $\mathcal{S} \approx 0.16$, only about half of the PIs can be expected to remain overlap-free. As the seeding density increases further, its effect on the number of overlap-free PIs begins to saturate. 
Finally, the scaling of the TCA and FRA can be approximated using a quadrature-like method. The relation of the TCA and FRA to the seeding density $\mathrm{S}$, provides crucial information on the expected fraction of a PI area which is not under influence of overlap and can be used in IPI, without restrictions.
Knowledge of expected values for key metrics, such as the average degree of an overlap cluster, the number of overlap-free PIs and the TCA, enables experimenters to design their setups (i.e. number of particles, defocus length, and aperture diameter) in a way that maintains PI overlap within acceptable limits on average. The derived scaling laws for each overlap metric serve as a look up table to design experiments. With the established scaling laws, critical seeding densities can be identified. These critical values enable the selection of appropriate combinations of PI sizes and particle numbers tailored to the specific requirements of a given experiment.
This approach helps to ensure that experimental data with appropriate PI overlap can be acquired in a quantitative manner instead of relying on an experimenters estimation from experience. 

\section*{Acknowledgments}

This work was supported by the Deutsche Forschungsgemeinschaft (DFG, German Research Foundation) via Project Grant KR4775/4-1 within the Research Unit FOR 5595 Archimedes (Oil-refrigerant multiphase flows in gaps with moving boundaries — Novel microscopic and macroscopic approaches for experiment, modeling, and simulation) — Project Number 510921053.
The authors furthermore thank Leister \textit{et al.},\cite{Leister.2021},  Pasch \textit{et al.},\cite{Pasch2024} and Pöppe\cite{poeppe2023} for kindly providing experimental raw images as used in this work.

\section*{Author contributions statement}

\textbf{CS:}
Conceptualization, 
Methodology,
Investigation, 
Software, 
Visualization, 
Data Curation, 
Formal Analysis, 
Validation,  
Writing – Original Draft Preparation, Review \& Editing; 
\textbf{JK:}
Conceptualization, 
Methodology,
Formal Analysis, 
Validation, 
Writing – Review \& Editing, 
Funding Acquisition, 
Supervision, 
Project Administration

\bibliography{main}

\clearpage
\section{Appendix: Particle Image Overlap in Single
Camera Defocusing Approaches without
Astigmatism}
\label{Appendix:Chap6_PI-Overlap}

\begin{figure}[htbp]
    \centering
    \begin{subfigure}[b]{0.3\linewidth}
        \centering
        \includegraphics[width=\linewidth]{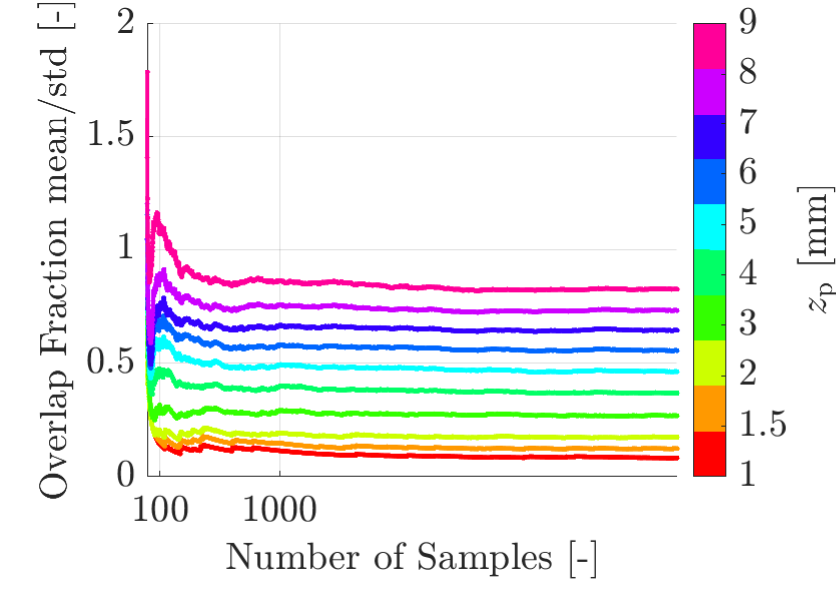}
        \caption{$N_\mathrm{P}=10$}
        \label{fig:Appendix_Chap6_StatConv-a}
    \end{subfigure}
    \hfill
    \begin{subfigure}[b]{0.3\linewidth}
        \centering
        \includegraphics[width=\linewidth]{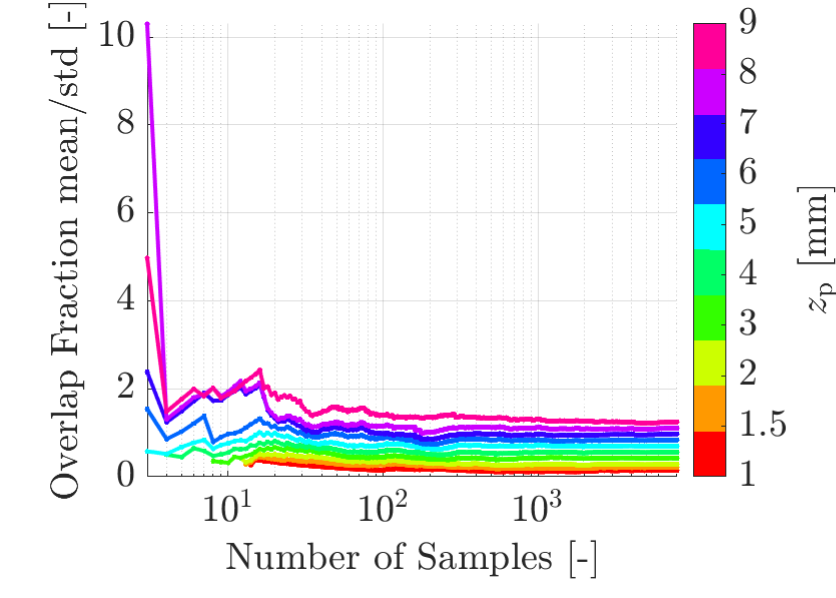}
        \caption{$N_\mathrm{P}=20$}
        \label{fig:Appendix_Chap6_StatConv-b}
    \end{subfigure}
    \hfill
    \begin{subfigure}[b]{0.3\linewidth}
        \centering
        \includegraphics[width=\linewidth]{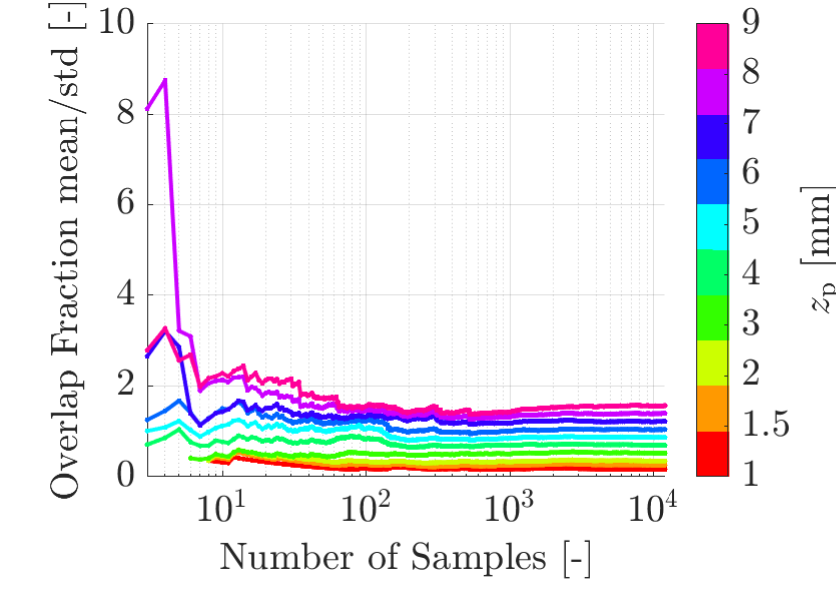}
        \caption{$N_\mathrm{P}=30$}
        \label{fig:Appendix_Chap6_StatConv-c}
    \end{subfigure}
    \hfill
    \begin{subfigure}[b]{0.3\linewidth}
        \centering
        \includegraphics[width=\linewidth]{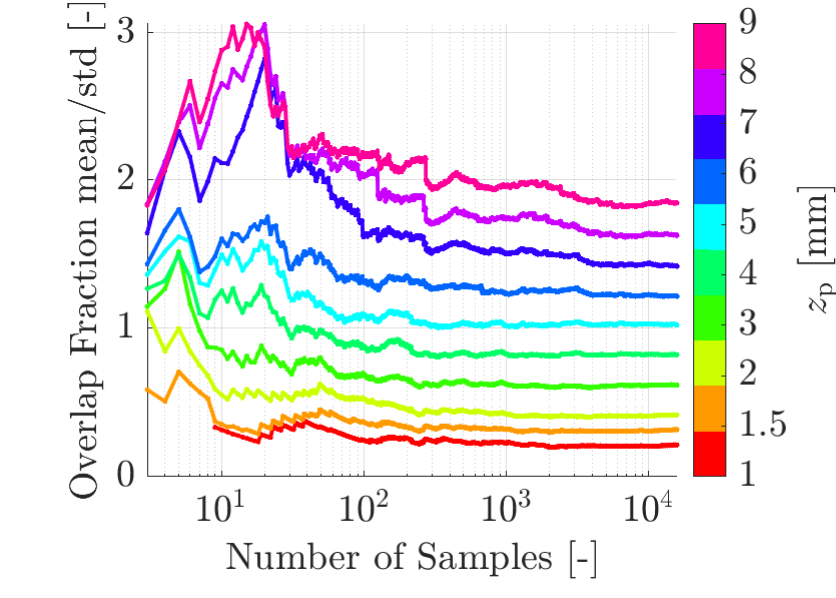}
        \caption{$N_\mathrm{P}=40$}
        \label{fig:Appendix_Chap6_StatConv-d}
    \end{subfigure}
    \hfill
    \begin{subfigure}[b]{0.3\linewidth}
        \centering
        \includegraphics[width=\linewidth]{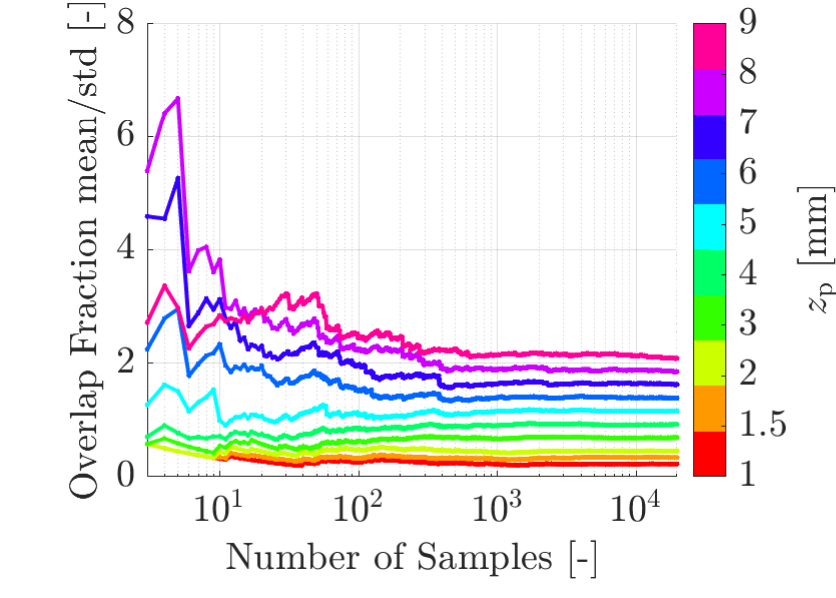}
        \caption{$N_\mathrm{P}=50$}
        \label{fig:Appendix_Chap6_StatConv-e}
    \end{subfigure}
     \hfill
    \begin{subfigure}[b]{0.3\linewidth}
        \centering
        \includegraphics[width=\linewidth]{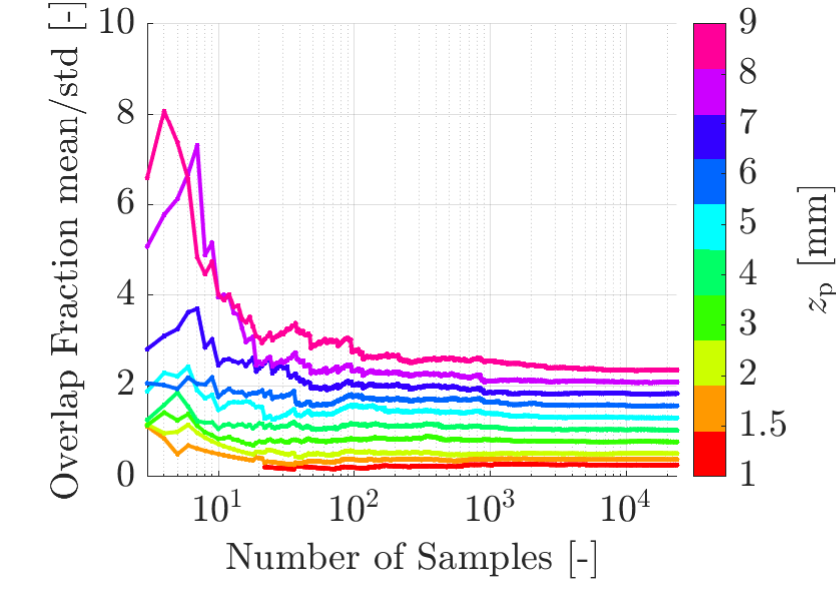}
        \caption{$N_\mathrm{P}=60$}
        \label{fig:Appendix_Chap6_StatConv-f}
    \end{subfigure}
    \caption{Test for statistical convergence by investigating the mean over the standard deviation of the overlap fraction for increasing dataset size. The $x$-axis is showing the number of PIs in the dataset. The depicted overlap fraction was computed by the discretized direct computation with an image of resolution 400\,px. The test were conducted for $N_\mathrm{P}=10,20,30,40,50,60$ particles per image (a),(b),(c),(d),(e) and (f) respectively, and for ten different PI sizes.}
    \label{fig:Appendix_Chap6_StatConv}
\end{figure}

The test for statistical convergence of the overlap fraction is shown in Fig.\,\ref{fig:Appendix_Chap6_StatConv}. The mean normalized by the standard deviation of the total covered area (TCA) is plotted for increased dataset size. It can be seen that a dataset size of 400 images with $N_\mathrm{P}=\{10,20,30,40,50,60\}$ PIs per image is sufficiently large to observe no changes in the TCA with further increasing dataset size.  

\end{document}